\documentclass{elsart}

\begin{document}

\begin{flushright}
Imperial/TP/01-2/5
\end{flushright}

\begin{frontmatter}

\title{Testing cosmological defect formation in the laboratory*}
\thanks{Text of an invited lecture at VORTEX II, Crete, September 2001}

\author{T.W.B. Kibble}

\address{Blackett Laboratory, Imperial College, London SW7 2BW, 
United Kingdom}

\begin{abstract}
{
Topological defects such as cosmic strings may have been formed at
early-universe phase transitions.  Direct tests of this idea are
impossible, but the mechanism can be elucidated by studying analogous
processes in low-temperature condensed-matter systems.  Experiments on
vortex formation in superfluid helium and in superconductors have so
far yielded somewhat confusing results.  I shall discuss their
possible interpretation.
}
\end{abstract}

\end{frontmatter}

\section{Introduction}

For quite a few years, my research has been mainly on the interface
between particle physics and cosmology.  So it may be quite surprising
that I am talking at a conference on Vortex Matter!  But there are good
reasons.  New connections have been forged in the last few years with
condensed matter physics --- connections in which vortices play a
central role, and that form the theme of the ESF Programme on {\em
Cosmology in the Laboratory}.  I want to explain how this came about. 
(I regret that Grisha Volovik, who co-chairs that Programme with me,
and who hoped to be here too, was unable to come.)

Our present understanding of fundamental particle physics leads us to
believe that very early in its history, the Universe underwent a
series of phase transitions.  The full symmetry of the underlying
theory is apparent only at extremely high energies.  As the Universe
expands and cools, the symmetry is progressively broken.  For example
the {\em electroweak\/} symmetry
\({\mathrm{SU}}(2)\times{\mathrm{U}}(1)\) is manifest only at energies
above a few hundred GeV; below that the symmetry is broken by the Higgs
mechanism to the \({\mathrm{U}}(1)\) of electromagnetic gauge
invariance.  At even higher energies, the symmetry may be larger
still.  There may be a {\em GUT\/} transition at an energy scale of
about \(10^{15}\) GeV, above which the strong, weak and electromagnetic
interactions are all combined in a {\em grand unified theory} with a
symmetry group such as \({\mathrm{SU}}(5)\) or \({\mathrm{SO}}(10)\). 
If, as is widely believed, the underlying theory is a {\em
superstring\/} or {\em M-theory}, the implied supersymmetry must be
broken at a transition scale of perhaps a few TeV.

But how can we test these ideas?  The energy scales are far too high
to be accessible to accelerator experiments.  The only place such
energies are found is the early Universe, in the first fraction of a
second after the Big Bang.  But of course the early Universe was
opaque; we have no direct observational access to it.  We have to
look for surviving traces of these very early events.  There are
various possibilities, but one in particular that I want to talk about. 
A common feature of symmetry-breaking phase transitions is the
formation of topological defects of one kind or another: monopoles,
cosmic strings or domain walls.  All these have analogues in
condensed-matter systems: hedgehogs, vortices or flux tubes, and
solitons.
 
Because of their topological stability, defects may have survived long
enough to be observable \cite{kib76,HK94,SV94}.  Indeed, monopoles and
domain walls, if formed, could have survived all too well, and already
be in conflict with observation, though there are possible ways out of
this.  Cosmic strings, however, are attractive from a cosmological
perspective.  They might help, for example, in explaining {\em
baryogenesis\/} --- creating the observed matter--antimatter asymmetry
of our Universe --- or {\em magnetogenesis} --- seeding the magnetic
field of galaxies.

To make definite predictions that can be tested against astronomical
observation, we need to answer several questions:
\begin{itemize}
\item{What {\em kinds\/} of defects can be formed in the hypothesized
phase transitions?}
\item{How {\em many\/} defects would be formed at the phase
transition?}
\item{How would they {\em evolve\/} as the Universe expands?}
\item{How would they {\em interact\/} with matter and radiation to
generate observable signatures?}
\end{itemize}
We have what we believe are good answers to these questions, based on
various assumptions and approximations.  But can we rely on them, when
our methods are untested?

Cosmology is an exciting subject, but it suffers from one major
shortcoming: we have no means of conducting controlled experiments, so
we cannot directly test our calculational methods.  It is here that the
analogy with condensed matter comes in.  The mathematical descriptions
of topological defects in particle-physics models and in
condensed-matter systems are often very similar.  The last of the
four questions above is a purely cosmological one, but for the others
analogies with condensed matter may be very instructive.  Our methods
of computing, for example, the number of defects formed at a
cosmological phase transition can also be applied to a transition in a
suitably chosen condensed-matter system, on which we can do real
experiments, an idea first suggested by Zurek \cite{zur85}.  Following
through this idea has led to some very exciting developments in
condensed-matter physics.

\section{Cosmic strings}

The simplest model in which strings or vortices are formed is
described by a complex scalar field \(\phi\) with \({\mathrm{U}}(1)\)
symmetry, \(\phi\to\phi\e^{{\mathrm{i}}\alpha}\).  This may be a global
symmetry or a local gauge symmetry.  In the latter case, \(\phi\)
interacts with a gauge potential \(A_\mu\) transforming according to
\(A_\mu\to A_\mu-{1\over e}\partial_\mu\alpha\).  The scalar potential
\(V\) must be \({\mathrm{U}}(1)\)-invariant.  If we choose
\[V(\phi)=\lambda(|\phi|^2-\eta^2)^2,\] 
(often called the {\em Mexican hat\/} potential), this is the Abelian
Higgs model, the relativistic analogue of the Ginzburg--Landau model. 
In this case, \(\phi=0\) is a maximum of the potential, so the
symmetry is broken.  There is a degenerate ground state, labelled by
the phase angle \(\alpha\):
\(\langle\phi\rangle=\eta\e^{{\mathrm{i}}\alpha}\).

This model (with or without the gauge field) exhibits a phase
transition.  Above a critical temperature \(T_{\mathrm{c}}\sim\eta\)
(I use units in which \(c=\hbar=k_\mathrm{B}=1\)) there are large
fluctuations in
\(\phi\) about a mean value of zero.  As the system is cooled through
the transition,
\(\phi\) falls into the trough of the potential and acquires a nonzero
average value.  In so doing it has to choose a phase \(\alpha\).  But
in a large system such as the Universe, there is no reason why this
choice should be the same everywhere; \(\alpha\) will vary randomly in
space.  Indeed, there can clearly be no correlation between the
directions of \(\alpha\) in regions beyond the causal horizon, which
have had no previous causal contact.

Now it may happen that if we traverse some large loop in space, the
value of \(\alpha\) will change by \(2\pi\) (or some multiple of
\(2\pi\)).  In that case, somewhere inside the loop \(\phi\) must go
through zero.  Indeed, it must vanish along a curve that threads its
way through the loop; this is the cosmic string.  It is the precise
analogue of the Abrikosov vortex in the Ginzburg--Landau model.

Because \(\phi\) has to climb over the central hump of the
potential, there is excess energy trapped on the string.  In the
local-symmetry case, this yields a {\em string tension\/} of order
\(\eta^2\).  When the symmetry is global, the energy per unit length
of an isolated string is logarithmically divergent, but in a real
system the divergence is cut off at the average string separation (or
the system size).  The strings are {\em topologically stable}.  Strings
will tend to shorten with time, under the effect of the string
tension, and small closed loops of string may shrink and disappear,
but they cannot break (except by a very exotic and unlikely process
involving the formation of a pair of black holes).  Because
\(\alpha\) is uncorrelated over large distances, the phase transition
will generate a random tangle of cosmic string.  The density of string
will tend to decrease with time, but some string is likely to survive a
long time, long enough to have observable consequences.

This \({\mathrm{U}}(1)\)  model is merely a simple example, but strings
also appear in many more realistic models, in particular in grand
unified theories.  Their dynamics and likely evolution are not usually
very different, however --- though there is one caveat: some strings
can support persistent currents carried by fermions trapped in zero
modes on the string, strongly influencing their behaviour.

\section{The Zurek predictions}

As the Universe cools through the relevant transition, we expect a
random tangle of string to be formed, with some characteristic scale
\(\xi_{\mathrm{str}}\).  In other words, in any randomly chosen volume
\(\xi_{\mathrm{str}}^3\), we expect to find on average a length
\(\xi_{\mathrm{str}}\) of string.  The question is: what is it that
determines this scale?  Of course, it must be related in some way to
the correlation length \(\xi\) of the scalar field.  But in the
neighbourhood of the transition, \(\xi\) is changing very rapidly. 
Indeed, at a second-order transition, the equilibrium correlation
length \(\xi_{\mathrm{eq}}\) goes to infinity at the transition
temperature \(T_{\mathrm{c}}\).  So this raises the question: at what
temperature, or what time, should we equate \(\xi_{\mathrm{str}}\) to
\(\xi\)?

Zurek \cite{zur93,zur96} has provided an answer to this question based
on a causality argument (see also \cite{kib80}).  When the system goes
through a real transition, at a finite rate, it is clear that \(\xi\)
cannot become infinite.  There is a maximum speed, \(c\), with which
correlations in the phase of the scalar field can propagate.  In the
relativistic case, this is the speed of light.  In a non-relativistic
system it is some characteristic speed of the system, for example the
speed of second sound in superfluid helium-4.  Correlations can never
extend beyond a finite range, determined by a balance between the
relaxation rate of the scalar field and the rate \(\dot T/T\) of the
transition.

Zurek's argument may be paraphrased thus: As the system cools, \(\xi\)
more or less keeps up, so long as it can do so, with the equilibrium
correlation length, \(\xi_{\mathrm{eq}}\).   But once
\(\d\xi_{\mathrm{eq}}/\d t\) becomes larger than \(c\), it can no
longer do so.  From then on, \(\xi\) does not change much until the
point after the transition when it again becomes equal to the
decreasing \(\xi_{\mathrm{eq}}\).  That is the time, now often called
the {\em Zurek time}, \(t_{\mathrm{Z}}\), at which we should identify
\(\xi_{\mathrm{str}}\) with \(\xi\).  An almost equivalent statement
(up to a factor of order 1) is that \(t_{\mathrm{Z}}\) is the time at
which correlations starting from zero at the transition and propagating
with speed \(c\), can reach the distance
\(\xi_{\mathrm{Z}}=\xi_{\mathrm{eq}}(t_{\mathrm{Z}})\).

This argument leads to a definite prediction for the defect density,
\(l\), that is, the average length of string per unit volume.  At
\(t_{\mathrm{Z}}\), we expect \(l(t_{\mathrm{Z}}) =
k/\xi_{\mathrm{Z}}^2\), where \(k\) is a constant roughly of order one. 
Numerical simulations \cite{LZ97,YZ98} suggest that \(k\) is actually
somewhat less than one, perhaps of order \(0.1\).

\section{Tests in Liquid Crystals and Helium-4}

Zurek \cite{zur85} originally suggested testing these predictions by
looking at a rapid quench in superfluid helium-4.  Starting at high
pressure in the normal phase just above the `lambda line', a rapid
redution of pressure takes the sample through the transition into the
superfluid phase.

The first experiments, however \cite{chu91,bow94}, were in thin samples
of nematic liquid crystals, where networks of defects were seen to be
formed when the system was rapidly cooled from the normal to the
nematic phase.  The nematic transition is first-order, so Zurek's
argument as given above is not directly applicable.  In that case,
what one expects \cite{kib76} is that \(\xi_{\mathrm{str}}\) is
approximately (a few times) the mean distance between nucleation
centres of bubbles of the new phase.  This indeed seems to be the case.

The helium-4 experiment, which provides a more direct test of Zurek's
predictions, has now been performed, twice, by a group in Lancaster,
using an apparatus comprising a small chamber filled with helium whose
sides were formed of bellows so that it could be rapidly expanded.  The
object of the exercise was to detect the presence of any vortices
formed during the transition by monitoring the absorption of second
sound, which is strongly attenuated by vortices.  The first experiment
\cite{hen94} did apparently show attentuation of the signal after the
quench, falling off exponentially with a timescale of the order of a
hundred milliseconds, as the vortices gradually disappeared, and
compatible in magnitude with Zurek's prediction for the number of
vortices.  But it was inconclusive for various reasons.  Firstly, the
detection equipment was swamped and unable to measure the attenuation
until about 50 ms {\em after\/} the quench, so the inferred
vortex density had to be extrapolated back to the Zurek time. 
Secondly, there were other possible sources of vorticity.  In
particular, vortices might have been nucleated by hydrodynamic effects
at the bellows, and a particular concern attached to the capillary
tube used to fill the chamber which was closed at its outer end, so
that fluid was injected into the chamber on each expansion.

For these reasons, the Lancaster group decided to build an improved
apparatus, in which the chamber was made much smoother, and in
particular the injection tube was closed at the point of entry into
the chamber, to minimize the risk of extraneous vortex formation. 
However, results with this improved apparatus \cite{dod98} rather
disappointingly failed to reveal any vortex formation.  I shall
discuss possible reasons for this later.

\section{Tests in Helium-3}

Meanwhile, experiments were performed using helium-3 at two different
laboratories, Helsinki and Grenoble (with collaboration from
Lancaster).  There are considerable advantages in using the lighter
isotope.  One is that the correlation length is much larger, say
40--100 nm as compared with less than 1 nm in \nuc{4}{He}.  This means
that a continuum description, of Ginzburg--Landau type, is a much
better approximation.  It also means that relatively speaking vortex
formation requires much more energy, so that extraneous vortex
formation is less likely.  But perhaps the most important difference
is that, because \nuc{3}{He} is an excellent neutron absorber, one can
perform a temperature- rather than pressure-driven quench.  Both
experiments made use of this technique.  The absorption of a neutron
in a sample in the superfluid \nuc{3}{He}-B phase heats up a small
region above the critical temperature.   It then cools rapidly, in a
period of about 1 \(\mu\)s, back through the transition into
the superfluid phase.  One expects the formation of a random tangle of
vortices during this process.

In other respects the experiments were very different.  The Helsinki
experiment \cite{ruu96} used a sample at a temperature not far below
the transition temperature, in a rotating cryostat.  If the rotation
speed \(v\) is less than some limit, no vortices are formed
spontaneously at the walls, so the superfluid component is actually
completely stationary, while the normal component is rotating with the
container.  The relative velocity between the two components means
that any vortex formed following neutron absorption will be subject to
the transverse {\em Magnus force}.  If a vortex loop is big enough and
appropriately oriented, the effect will be to expand it until it meets
the walls of the container, after which the ends migrate to the top
and bottom surfaces, and the vortex joins a central cluster parallel
to the rotation axis.

Detection of these captured vortices is made possible by another of the
key features of \nuc{3}{He}, its non-zero nuclear spin, which makes it
possible to use nuclear magnetic resonance.  In the NMR trace one can
actually see each individual vortex joining this central cluster.  It
turns out that the Zurek scenario leads to a very simple prediction
for the dependence of the number of vortices captured per neutron
event on \(v\).  There is a critical velocity \(v_{\mathrm{c}}\) below
which no vortices are captured.  When \(v>v_{\mathrm{c}}\), 
\[n=\gamma\left\{\left(v\over v_{\mathrm{c}}\right)^3-1\right\},\]
where \(\gamma\) is a constant.  Remarkably the entire dependence on
the pressure, the bulk temperature and the magnetic field is contained
in the single parameter \(v_{\mathrm{c}}\).  The results clearly
confirm this: plotted against \(v^3\), they fit well to straight lines,
with a common intercept at \(-\gamma\).  Moreover the dependence of
\(v_{\mathrm{c}}\) on the bulk temperature also fits the prediction.

The Grenoble experiment \cite{bau96} was complementary.  Essentially it
involved calorimetry.  They used a sample at a much lower temperature,
but in a non-rotating container, and sought to measure the energy
release every time a neutron was absorbed.  The absorption reaction is
\[n+\mbox{\nuc{3}{He}}\to p+\mbox{\nuc{3}{H}}+764\;\mathrm{keV}.\]
The total energy released in the form of quasiparticles was measured,
and found to peak in the range 575--650 keV, depending on the
pressure.  Some energy, around 50 keV, is also released in the form of
ultraviolet radiation.  However, there is a very clear overall
deficit, which is attributed to vortices.  It is very difficult to
think of any other mechanism of energy loss.  Moreover, the magnitude
and the dependence on pressure are entirely consistent with Zurek's
prediction.

\section{Tests in Superconductors}

From the point of view of analogy with cosmological phase transitions,
tests in superconductors are particularly important, because they
provide an example of symmetry breaking in a {\em local gauge theory}.

At least two such experiments have in fact been performed, by a group
at Technion.  In the first \cite{car99}, they used a 1 cm\(^2\) thin
film of the high-temperature superconductor, YBCO.  The film was
heated to above \(T_{\mathrm{c}}\) by optical irradiation, and then
allowed to cool back through  the transition.  To determine the number
of flux tubes formed during this process, the experimenters measured
the total flux using a SQUID	close to the sample, but separated from it
by a mylar sheet.  To be precise, this process does not measure the
number of fluxons, but only the net total flux, the number of fluxons
minus the number of antifluxons.  A straightforward application of
Zurek's technique would lead to an estimate for the net flux of about
100, whereas Carmi and Polturak's results \cite{car99} were consistent
with zero, with an error of about 10.  Moreover, they argue that
because of the different behaviour of a gauge theory, the actual
prediction should be higher, around 10\(^4\), in which case there is a
very clear contradiction.  However, this prediction does not seem to be
on very firm ground.

The same group have done another, similar experiment involving a loop
of semiconductor with a large number of Josephson junctions in
series.  In that case they do find definite evidence \cite{car00} of
the trapping of magnetic flux during the cooling process.  Here the loop
becomes superconducting before being linked by the junctions, so one
might expect the phases to be random.  The experiments actually show a
{\em larger} flux than predicted on that basis, suggesting that the
phase differences are not uniformly distributed, though it is not
entirely clear why.

Experiments have been suggested \cite{kav00} using an annular
Jospehson junction, in which cooling can lead to the trapping of
radial flux lines.  A group in Salerno has already done some
experiments with an apparatus of this kind, but the experimental
parameters were not optimized for this particular type of measurement: 
the predicted average number of trapped quanta is less than unity. 
Future experiments along these lines should yield interesting results.  

\section{Theoretical Interpretation}

Zurek's predictions of the numbers of defects formed are based on an
admittedly rather crude argument, and it is certainly not surprising
to find that it needs modification.

At first sight, the difference between the behaviour of helium-3 and
helium-4 may seem surprising, though in fact theoretically they are
very different types of system.  There is a critical region below
\(T_{\mathrm{c}}\) but above the so-called {\em Ginzburg
temperature\/}, within which there is a large population of transient
thermal loops.  This region is much wider in helium-4 than in
helium-3.  Work by Karra and Rivers \cite{kar98} has shown that this
has a marked effect on the Zurek predictions.  On the other hand, there
may be a more immediate reason for the failure of the helium-4
experiment to detect any vortices: they may simply not last long enough
to have been seen.  Vorticity does decay with time, with a lifetime
that is not well known.  The authors \cite{hen94} estimated the
lifetime on the basis of measurements of decay of turbulent vorticity,
but this may well be inapplicable \cite{riv00,riv01}.  It would be very
desirable to measure the vorticity at a much earlier time after the
transition.  The situation should be clarified by future experiments
\cite{hen00}.

Understanding the results in superconductors also presents a
challenge.  It is puzzling that the experiment with a thin film of
YBCO saw nothing.  One of the problems is that there is much more
theoretical uncertainty about the predicted defect density in the case
of a local gauge theory.  There is a competing mechanism for the
formation of defects, namely the thermal fluctuations in the magnetic
field.  Hindmarsh and Rajantie \cite{hin01} have shown that that this
mechanism yields a very different density and distribution of
defects.  It is unclear which mechanism will dominate.  Neither
mechanism seems to support the estimate that led to the conclusion of
a disagreement by a factor of 1000 in the experiment with a thin
superconducting film, but there remains {\em at least} a factor of 10
that is hard to explain. 

The Josephson junction results seem to be positive, but if anything
show too large a trapped flux.  Why there should be this difference
between the two experiments with superconductors is again something of
a puzzle.  More experiments along these lines too are very desirable.

\section{Conclusions}

The results so far are distinctly confusing.  By far the best evidence
in favour of the Zurek scenario comes from the experiments in
helium-3, both of which yielded positive results that are hard to
interpret in any other way.  Results in nematic liquid crystals are
also positive, but not a direct test of the Zurek scenario.
In superconductors, it is puzzling that the experiment with a thin YBCO
film saw no fluxons, while more than expected were seen in the
Josephson-junction array.

It does seem that vortex formation in a rapid transition is now well
established, but there is still a lot of work to do to understand the
details.

\section*{Acknowledgements}

I am very grateful to the organziers of this meeting for inviting me
to speak.  The work reported here was supported by the European
Science Foundation under the COSLAB Programme (Cosmology in the
Laboratory).

\end{document}